\documentclass[letter]{aa} % for the letters 
\usepackage{graphicx}
\usepackage{txfonts}
\usepackage{ctable}
\usepackage{natbib}
\usepackage{tabularx}
\def\bpcru {\hbox{BP Cru}}
\def\gx {\hbox{GX~301-2}}

\def\bex {\hbox{Be$/$X-ray}}
\def\gx {\hbox{GX301-2}}
\def\wr {\hbox{Wray~977}}

\def\mjyb {\hbox{mJy\,beam$^{-1}$}}
\def\msun {\hbox{\rm{M}$_{\odot}$}}
\def\rsun {\hbox{\rm{R}$_{\odot}$}}

\begin{document}
   \title{Radio emission from the high-mass X-ray
     binary \bpcru} 

   \subtitle{First detection}

   \author{M. Pestalozzi
          \inst{1}
          \and
          U. Torkelsson\inst{1}
          \and
          G. Hobbs\inst{2}
          \and
          \'A. R. L\'opez-S\'anchez\inst{2}
          }
   \offprints{M. Pestalozzi}

   \institute{Department of Physics, University of Gothenburg,
              412 96 G\"oteborg, Sweden\\
             \email{torkel,michele.pestalozzi@physics.gu.se}
             \and
             Australia Telescope National Facility, PO Box 76, Epping,
              NSW 1710, Australia \\ 
              \email{George.Hobbs,Angel.Lopez-Sanchez@csiro.au}
             }

   \date{To be submitted to A\&A}

% \abstract{}{}{}{}{} 
% 5 {} token are mandatory

  \abstract
   {\bpcru\ is a well known high-mass X-ray binary composed of a late B
     hypergiant (\wr) and a neutron star, also observed as the 
     X-ray pulsar \gx. No information about emission from \bpcru\ in
     other bands than X-rays and optical has been reported to date in
     the literature, though massive X-ray binaries containing black
     holes can have radio emission from a jet.}
    {In order to assess the presence of a radio jet, we searched for
      radio emission towards \bpcru\ using the 
      Australia Compact Array Telescope during a survey for radio
      emission from Be/X-ray transients.}
   {We probed the 41.5\,d orbit of \bpcru\ with the
     Australia Telescope Compact Array not only close to periastron but
     also close to apastron.}
   {\bpcru\ was clearly detected in our data on 4, possibly 6, of 12
     occasions at 4.8 and 8.6\,GHz. Our data suggest that the spectral
     index of the radio emission is modulated either by the X-ray flux
     or the orbital phase of the system. }
   {We propose that the radio emission of \bpcru\ probably arises from
     two components: a persistent component, coming from the mass
     donor \wr, and a periodic component connected to the accretion onto
     the neutron star, possibly coming from a (weak and
     short lived) jet. }


   \maketitle
%
%________________________________________________________________

\section{Introduction}

The majority of High-Mass X-ray Binaries (HMXB) are classified as
\bex\ transients, i.e. the compact object in these systems is the 
companion of a Be-star. Some \bex\ transients display X-ray
outbursts at most of periastron passages as the result of mass
transfer from the Be-star. Another important group of HMXBs contains
systems were the mass donor is a OB super- or hypergiant, that is
losing mass  through a massive stellar wind, from which the compact
object is accreting.

The wind accreting X-ray pulsar \gx\ (also known as 4U~1223-62) is part
of the well known high-mass X-ray binary \bpcru. This system is
peculiar in several aspects (Table~\ref{tab:wray977}): it contains the
most massive star known in these systems and one among the slowest
pulsars.

The orbital period of \bpcru\ is $P_{orb}$ = 41.5\,d, and
the X-ray flares occur about two days {\it before}
periastron (\citealt{wat82}).  
In addition \bpcru\ shows a weaker X-ray flare close to apastron. 
The fact that the strong X-ray outbursts occur
close to periastron makes it similar to the \bex\ transients.  
Models of fast and dense stellar winds from
the primary have difficulties in reproducing the correlations of the weak
and strong X-ray outbursts with the orbital phases of \bpcru.
Among the alternatives are the existence
of a circumstellar disc around \wr\ through which the neutron star
passes at every periastron, an enhancement of the stellar wind close
to periastron (\citealt{pra95}), as well as a spiral-type tidal stream
of gas trailing the mass donor's rotation (\citealt{lea08}), that can
produce a transient accretion disc around the neutron star, which
explains the episodes of rapid spin-up of the X-ray pulsar
(\citealt{koh97}).

While \bex\ transients are easily observed in the X-rays and some of them 
have known optical counterparts, very few radio observations of these
objects have been reported in the literature. The radio surveys of
X-ray binaries by \citet{dul79} and \citet{nel88} report only upper
limits for 7 \bex\ transients, and one positive detection of 14\,mJy at
2 cm (April 1978) for A 1118-61 \citep{dul79}. For this source they
found only upper limits on the 3\,mJy level at 7 other epochs. The
orbital phases of the system during these observations are
unclear. The main problem for the observation of these systems is the
transient nature of the outbursts, and in some cases the coarse X-ray
positions. 

Some other X-ray binaries (as e.g. microquasars), have relativistic
jets that produce synchrotron emission at radio wavelengths. In
general microquasars are black 
hole candidates, but in a few cases the compact object might rather be
a neutron star.  The neutron star systems that are emitting radio
waves tend to have high luminosities and weakly magnetised neutron
stars. On the other hand, X-ray pulsars that have strongly
magnetised neutron stars but are less luminous, are as a rule not
detected at radio wavelengths. 

Outbursts in microquasars are characterised by an X-ray outburst close
to periastron followed by the formation of a radio jet at a later
time, when the X-ray spectrum indicates that the inner part of the
accretion disc has vanished (\citealt{har95}). The radio outburst can
therefore be delayed by up to some 10 days relative to the X-ray
outburst. 
 
With this in mind, we devised an observational programme to search for
radio emission from \bex\ transients and other HMXBs that have
semi-regular X-ray outbursts close to periastron. Our strategy was to
distribute "snap-shot" observations to cover the orbital phases
between periastron and apastron. 

\begin{ctable}[
notespar,
caption={Characteristics of the \bpcru\ system.},
captionskip = 0ex,
label={tab:wray977},
width=\columnwidth,
center
]{lXlX}{
\tnote[a]{\citealt{kap06}}
\tnote[b]{\citealt{kre04}}
\tnote[c]{\citealt{whi76}}
}{ \FL
\multicolumn{2}{l}{\bpcru} 
\ML
RA(J2000) 12:26:37.6 & DEC(J2000) $-$62:46:14 \\
Orbital period & 41.5\,d \\
\FL
\multicolumn{2}{l}{\wr\tmark[a]}
\ML
Mass & 39 - 68\,\msun \\
Radius & 62\,\rsun \\
$T_{eff}$ & 18100\,K \\
Luminosity & $5\times 10^{5}\,L_{\odot}$ \\
Mass-loss rate & $10^{-5}$\,\msun\,yr$^{-1}$ \\
Wind velocity & 305\,km\,s$^{-1}$ \\
\FL
\multicolumn{2}{l}{\gx}
\ML
Mass\tmark[a] & 1.85\,\msun\\
Spin period\tmark[c] & 696\,s \\
Surface magnetic field\tmark[b] & $4\times 10^{8}$\,T 
\LL}
\end{ctable}

In this letter we present the first results of our observational
campaign, the detection of radio continuum
emission in the centimetre band at two frequencies (4.8 and 8.6\,GHz)
from the \bpcru\ system.  

%__________________________________________________________________

\section{Observations and data reduction}
\label{sec:obs}

\begin{figure}
\centering
\includegraphics[width=7.0cm]{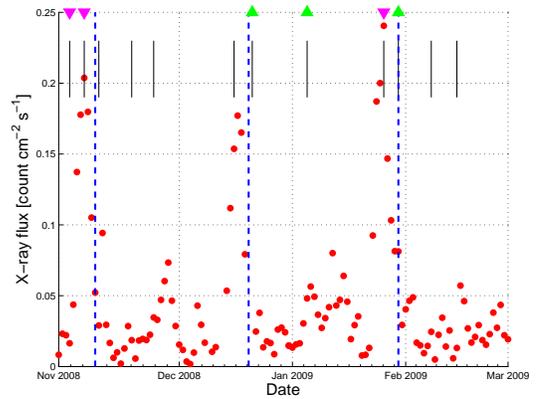}
\caption{X-ray lightcurve (Swift/BAT, 15 - 150 keV) of \bpcru\ in
  the period of our observing  
  campaign (red dots). Vertical black lines
  indicate the epochs of our radio observations, and the dashed blue lines 
the periastron passages. Green triangles peak up indicate a positive
spectral index, magenta triangles peak down indicate negative spectral
index.} 
\label{fig:obs_bpcru}
\end{figure}

Observations of \bpcru\ were done at 12 epochs for a total of
approximately 20 hours of observations, 13 of which on target. The
observation epochs were spread between November 2008 and February
2009, covering almost 
four complete orbits. Every epoch consisted of about 1.5 hours of data
recording, and this time was divided between flux and phase
calibrators and the target. The original strategy was to spread our
observations over different orbital phases, from shortly before
periastron to close to apastron. Our proposed schedule was then folded
with the telescope availability to give the final schedule summarised
in Table~\ref{tab:epochs} and graphically presented in
Fig.~\ref{fig:obs_bpcru}. 

\begin{table}
\begin{minipage}[t]{\columnwidth}
\caption{Details of the observations presented in this paper.}
\label{tab:epochs}
\centering
\renewcommand{\footnoterule}{} 
\begin{tabular}{ccccc}
\hline\hline
Epoch & Date (UT)\footnote{Times indicate the beginning of every
  observation} & T-on [min]\footnote{Effective time spent on-source} & 
Phase\footnote{Days before ($-$) or after (+) periastron. 
Periastron passage occurred at 2008-11-08, 2008-12-20, and 2009-01-30 (see
  \citealt{koh97}).} & Array\footnote{ATCA configuration for the
  observations. When including all antennae, the {\it uv}-coverage from
  the '6'-configurations is 
  regular, from the '1.5'-configurations slightly bimodal and from the EW-
  and 750-configurations (compact configurations) it is strongly
  bimodal.}\\
\hline
1 & 2008-11-01, 2212 & 36 & $-$7 & 6A \\
2 & 2008-11-05, 1708 & 54 & $-$3 & 6A \\
3 & 2008-11-09, 0443 & 63 & +1 & 6A \\
4 & 2008-11-18, 0314 & 45 & +10 & EW367 \\
5 & 2008-11-24, 2055 & 45 & +16 & EW367 \\
6 & 2008-12-16, 1936 & 54 & $-$4 & 750B \\
7 & 2008-12-21, 1840 & 54 & +1 & 6C \\
8 & 2009-01-05, 1719 & 153 & +16 & 6C \\
9 & 2009-01-26, 1603 & 63 & $-$4 & 1.5C \\
10& 2009-01-30, 1415 & 72 & +0 & 1.5C \\
11& 2009-02-08, 1936 & 72 & +9 & EW352 \\
12& 2009-02-15, 1122 & 54 & +16 & EW352 \\
\hline
\end{tabular}
\end{minipage}
\end{table}

We used the Australia Telescope Compact Array (ATCA) in the 4.8 and
8.6\,GHz continuum bands in two linear polarisations. Data were
recorded at the largest bandwidth, 128\,MHz, spread over 33 channels 
per polarisation and frequency. All four polarisation products were
recorded (XX, YY, XY, YX) in order to retrieve information about the
level of polarisation of a detection. The time in every observation
block was divided between a 
phase calibrator (1148-671), and the target, alternating 2 minutes
scans on the calibrator with 6-8 minutes on the target. Additionally, at
the beginning of each run, a bandpass calibrator was  observed for some
5-10 minutes. Since the project was a detection
experiment, no requirements were made as to which configuration ATCA
had to be in at the different epochs. This had the advantage of greatly
improving the scheduling flexibility but the disadvantage of having to
reduce data in all kinds of configurations. In this respect, data at
epochs 4, 5, 6, 11 and 12 (see Table~\ref{tab:epochs}) turned out to
be less useful for our program because they were recorded in the most
compact ATCA configurations (see Table~\ref{tab:epochs}) and are hence
heavily biased by the Galactic emission, as \bpcru\ lies in the plane
of the Galaxy. The 1\,$\sigma$ rms level in those maps was on average
$\approx 0.3$\,mJy (i.e. a detection limit of $\approx$0.9\,mJy) at both
frequencies. To prevent loss of sensitivity on  
target due to possible artefacts, the ATCA array was pointed
$\pm$10 arcsec away from \bpcru. Finally, most of the observations
were done in remote mode, making use of the exceptionally functional
software available for ATCA. 

%It turned out that every second channel in our observations was heavily
%affected by a DC offset giving a strong artefact at the image centre. This
%problem could be solved by reading only the odd channels at the
%centre of the band into the reduction software. This trick does not reduce the
%sensitivity in the data, as the 33 original channels are not
%completely independent\footnote{See the MIRIAD
%  documentation on the ATCA website.}. 

Bandpass, flux and phase calibration were performed using
standard procedures in AIPS\footnote{http://www.aips.nrao.edu/} with a
ParselTongue\footnote{http://www.radionet-eu.org/rnwiki/ParselTongue}
interface to allow quick and equal reduction 
of many observational epochs, as well as in 
MIRIAD\footnote{http://www.atnf.csiro.au/computing/software/miriad/}. The
bandpass calibrator was 0823-500 (1934-638 for epochs 3 and 4). The
flux scale was set by the Stokes I fluxes at 4.8 and 8.6\,GHz of the
phase calibrator 1148-671, i.e. 1.43 and 0.92\,Jy, respectively. All
solutions obtained toward the calibrators were applied 
to the target and images were made using different restoring beam
sizes to reflect the different array configurations of ATCA in the
different epochs. Our short observations gave elongated beam sizes, the
average beam sizes for the epochs with detections (see
Table~\ref{tab:det}) 
were $15\times 1.5$ and $10\times 0.8$\,arcsec at 4.8 and 8.6\,GHz,
respectively. The integration time on target varied between 36 
and 153 minutes (see Table~\ref{tab:epochs}), which allowed us to reach
rms levels in the final maps between 0.16 and 0.09\,\mjyb ~at
4.8\,GHz and between 0.17 and 0.09\,\mjyb ~at 8.6\,GHz. 
%AIPS and MIRIAD are
%in principle equivalent, but MIRIAD is better equipped to extract
%polarisation information from data recorded with linear orthogonal
%feeds. 
Even epoch 8, our longest observation of \bpcru, was too short
and covered a too short range of hour angles to allow a safe
correction for polarisation leakage. What we report here in terms of
polarisation is a crude estimate done for only one epoch. 

\begin{ctable}[
caption={Details of the detections and upper limits of the
  non-detections of \bpcru ~ordered in orbital phase.}, 
label={tab:det},
center,
width=\columnwidth
]{cccc}{
\tnote[a]{Phase indicates ``days from periastron''.}
\tnote[b]{Numbers in parenthesis indicate the rms value in the maps.}
\tnote[c]{Spectral index, $\alpha=\log(F_{8.6}/F_{4.8}) /
  \log(\nu_{8.6} / \nu_{4.8})$}
\tnote[d]{Orbital phase also observed in epoch 6 with a compact
  configuration, and hence not considered useful for this project.}
\tnote[e]{Orbital phase also observed in epoch 3 without
  detection. See text.}
\tnote[f]{Note that for this epoch Stokes Q at 8.6\,GHz is 0.38\,mJy,
  while Stokes V and U were not detected. No polarisation signal was
  detected at 4.8 GHz.}
\tnote[g]{Maximum and minimum values of the spectral index when
  adding/subtracting 1$\sigma$ rms to the listed flux.}
\tnote[h]{As mentioned in the text, we do not detect \bpcru\ at this epoch.}
}{\FL
 Epoch (phase)\tmark[a] & \multicolumn{2}{c}{Flux [mJy]\tmark[b]} &
 $\alpha$\tmark[c]\tmark[g] \\ 
 & 4.8\,GHz & 8.6\,GHz &
\ML
 1($-$7) & 1.02 (0.17) & 0.82 (0.18) & $-0.38^{+0.28}_{-1.10}$ \\
 & & & \\
% 6($-$4) & (0.38) & (0.19) & \\
{\it 9($-$4)}\tmark[d] & {\it 1.01 (0.16)} & {\it 0.51 (0.13)} & {\it
  {\it $-$1.17$^{-0.49}_{-1.93}$}} \\ 
 & & & \\
 2($-$3) & 0.75 (0.14) & $<0.48$ (0.16) & $-0.88^{+0.0}$ \\
 & & & \\
10(+0) & 0.70 (0.14) & 0.72 (0.13) & $+0.05^{+0.72}_{-0.61}$ \\
 & & & \\
 3(+1)\tmark[h] & (0.18) & (0.15) & \\
 & & & \\
{\it 7(+1)}\tmark[e] & {\it 0.67 (0.13)} & {\it 0.91 (0.14)} & {\it
  {\it +0.53$^{+1.14}_{-0.07}$}} \\ 
 & & & \\
%11(+9) & (0.40) & (0.25) & \\
% 4(+10) & (0.37) & (0.46) & \\
% 5(+16) & (0.29) & (0.17) & \\
 8(+16)\tmark[f] & 0.51 (0.09) & 0.63 (0.09) & $+0.36^{+0.92}_{-0.18}$ \\ \\
%12(+16) & (0.47) & (0.28) & \\
\FL }
\end{ctable}

\begin{ctable}[
caption={Characteristics of the objects detected around
  \bpcru. Coordinates and fluxes refer to epoch 8.},
label={tab:other_det},
center,notespar
]{ccccl}{
\tnote[a]{within 7.5''},
\tnote[b]{within 9.5''},
\tnote[c]{within 1.2''}
}{\FL
 & RA(J2000) & DEC(J2000) & Flux  & Remarks \\
 & & & 4.8/8.6 [mJy] &  
\ML
A & 12:26:23.9 & $-$62:45:58.75 & 1.53 / 1.10 &
2MASS\tmark[a] \\
B & 12:26:32.9 & $-$62:44:55:59 & 1.16 / 0.96 &
2MASS\tmark[b] \\
C & 12:26:28.6 & $-$62:46:17.58 & 0.70 / 0.67 & GSC2.3\tmark[c] 
\LL}
\end{ctable}

%================================================
\section{Results}
\label{sec:res}

A summary of the detections of \bpcru\ is presented in
Table~\ref{tab:det}. \bpcru\ was clearly detected at 4 epochs,
corresponding to 4 different orbital phases: two
before periastron, one after periastron and one at periastron. The
clearest detection was obtained in our longest integration on-source
at epoch 8, 16 days after periastron. The cleaned map at 8.6\,GHz is
presented in Fig.~2. All other detections are 
significant to at least 5$\sigma$\,rms. We report also on two more
possible detections, at epochs 7 and 9. The uncertainty of the
detections comes from the fact that the same orbital phase was
observed twice, with a detection in only one of the two
occasions. Epochs 6 and 9 correspond to 4 days before periastron, but
only epoch 9 was useful (see Sect.~\ref{sec:obs}); epochs 3 and 7 probed
\bpcru\ one day after periastron but a detection could be seen only in
epoch 7. Apart from these uncertainties, the most striking aspect
here is that \bpcru\ seems to emit in the radio regime regardless of
the orbital phase and especially when it was least expected,
i.e. close to apastron.

Another important observation from Tab.~\ref{tab:det} is that $\alpha$
appears to change from being clearly negative right before periastron
to being weakly positive after periastron. As visible from the minimum
and maximum values listed 
in Table~\ref{tab:det} $\alpha$ seems to be plausibly positive {\it
  at} and {\it after} periastron, while it is possibly negative before
that. Notice that \bpcru\ was not detected at 8.6\,GHz in epoch 2 and we 
therefore consider the 3$\sigma$\,rms level as an upper limit for the
flux. Furthermore it is important to notice that the epochs with
possibly negative spectral index correlate with the X-ray 
outburst, which in its turn is known to occur some days before
periastron (\citealt{wat82}). Possible explanations of this
behaviour are discussed in the next section. 

Finally, from polarisation analysis of the longest observation of
\bpcru\ we notice that its radio emission is probably linearly
polarised, as Stokes Q $\ne 0$ and Stokes V=U=0. This result is to be
taken with caution because our 
observation at that time was shorter than three hours, barely enough to
correct for polarisation leakage between the ATCA orthogonal
feeds. Dedicated observations are necessary to retrieve more reliable
information about the polarisation of the radio emission from \bpcru. 

\begin{figure}
\label{fig:maps}
\centering
\includegraphics[width=\columnwidth]{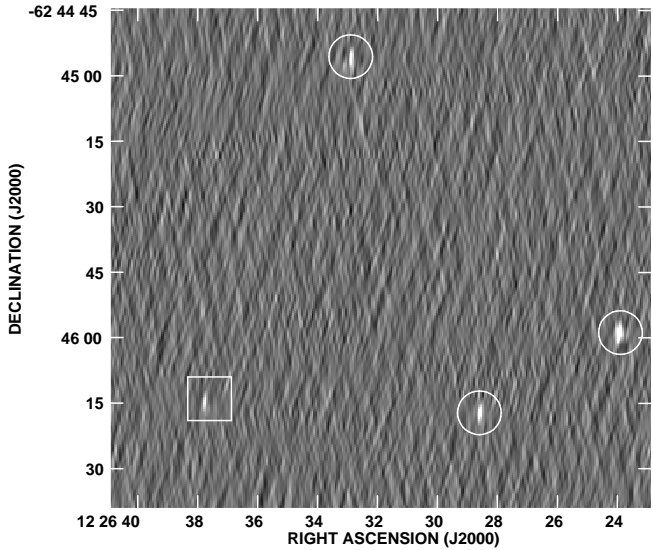}
\caption{Stokes I maps of the data taken in
  epoch 8, at 8.6\,GHz. The beam size for
  this epoch was $3.4\times 0.7$\,arcsec. The box
  indicates \bpcru, the circles indicate sources A (to the right),
  B (top in the figure) and C (lowest in the figure) as
  in Table~\ref{tab:other_det}.} 
\end{figure}

Several other sources were detected in the ATCA field, these are
listed in Table~\ref{tab:other_det}. Sources A and B where detected at all
useful epochs. Source A appears to lie some 7.5\,arcsec away from a
2MASS catalogued object, 12262412-6246060, with {\it JHK} magnitudes 15.83,
15.02, 18.3, respectively. Source B lies some 9.5\,arcsec from a 2MASS
catalogued source, 12263253-6244464,  with {\it JHK} magnitudes 15.67,
14.20, 13.60, respectively. Source C lies at about 1.2\,arcsec from an
object listed in the Guide Star Catalogue (GSC)2.3 and has a {\it V}
magnitude of 16.71. The closest 2MASS object to C is 12262821-6246162
at 2.8\,arcsec distance, and having {\it JHK} magnitudes 15.03, 14.64,
14.42, respectively. Taking into account the orientation of the array,
and considering the large beam sizes in our observations, our
positions of sources A, B and C are consistent with those of the catalogued 
objects mentioned above. The flux densities of these sources
  remained constant within $\approx$15\%. A further discussion of
these sources though goes beyond the scope of this paper. 

%__________________________________________________________________
\section{Discussion}
\label{sec:dis}

A spherical stellar wind from \wr\ is expected to generate continuum
fluxes of 1.9 and 2.6\,mJy at 4.8 and 8.6\,GHz, respectively, assuming
a distance of 3\,kpc for \wr\ (\citealt{pan75,wri75,scu98}), which is more
than a factor of two higher than what we detect.  There are several
ways in which the predicted fluxes can be reconciled with the observed
fluxes. Firstly, there are large uncertainties in the distance to
\bpcru, and a distance of 4-5\,kpc, sufficient to decrease our
observed fluxes with a factor of two, is still perfectly compatible with
the most recent estimates (\citealt{kap06}). Secondly it is not clear
that \wr\ is emitting a sufficient number of Lyman photons to keep the
wind completely ionised. In fact, using \citet{scu98} we estimate that
\wr\ produces some 3-5 times less photons than what is required to
keep the entire wind ionised. Also, it is to be noticed that the
  radio flux from a stellar wind can vary by a factor two
  (\citealt{scu98}).

Typically, the radio emission from the stellar wind of an early
  type star has a positive spectral index (\citealt{scu98}).  We
  suggest therefore that the radio emission at epochs 7 and 8 is due
  to the stellar wind from \wr.  Consequently, the increased radio
  flux at 4.8\,GHz at the time of the X-ray outburst is probably due
  to a transient, non-thermal component with a negative spectral index.
  There is not enough data, to tell whether this component is the most
  strongly correlated with the orbital phase or the timing of the
  X-ray outburst.

We propose that the transient non-thermal component appearing during the
X-ray outburst originates from a weak and short lived (relativistic?)
jet, as in a microquasar. To test this hypothesis, we isolated the
transient component by subtracting the flux from 
epoch 8 from that at epoch 1. The transient component has then fluxes
of 0.51 and 0.19\,mJy at 4.8 and 8.6\,GHz, and a spectral index
$\alpha=-1.7^{+0.1}_{-7.0}$. This suggests a somewhat steeper  index
than is usually seen in microquasars (e.g. \citealt{mcli09,tru08}, in
which $\alpha > -1$, but the uncertainties 
are so large that no firm conclusions can be drawn.  We also compare
\bpcru\ with the microquasar GRS~1915+105.  During an extended period in
2008 Swift/BAT registered an X-ray flux of 0.1\,ph\,cm$^{-2}$\,s$^{-1}$
from GRS~1915+105, and at the same time its radio flux at 4.8 GHz
remained above 100\,mJy (\citealt{tru08}).  The X-ray flux from
\bpcru\ at the peak of an outburst is even higher, 0.2
\,ph\,cm$^{-2}$\,s$^{-1}$, but the flux of its transient radio component
is at least 200 times fainter than that of GRS 1915+105.

A problem with this interpretation is that the detection at epoch 1
occurs so close to the onset of the X-ray outburst that one might
question whether a jet can have formed already at that time.  An
alternative source of the non-thermal electrons would be a shock that
forms in the stellar wind as the neutron star approaches \wr\ and
distorts the stellar wind through its gravitational field. 

In the future one should follow the evolution of the radio flux from
\bpcru\ over all orbital phases, and every observation should be
sufficiently long to allow the study of the polarisation of the radio
emission. This is readily possible using the Compact Array Broadband
Backend (CABB), newly installed at ATCA. Its wide band increases the
overall sensitivity by a factor 4, ensuring the detection of
\bpcru\ at any orbital phase. Also, with sufficiently long
integrations to give good signal-to-noise ratios, it will be possible
to obtain a full spectrum in two bands with ideally one point every
250\,MHz.

%__________________________________________________________________
\section{Summary and Conclusions}
\label{sec:con}

We summarise our findings as follows:

\begin{itemize}
\item We have presented the first detection of the high-mass X-ray binary
\bpcru\ in centimetre radio continuum at 4.8 and 8.6\,GHz. The 
source has been detected several times at different orbital phases,
indicating that \bpcru\ emits radio waves regardless of the orbital phase;

\item The radio emission from \bpcru\ shows a varying spectral index,
  suggesting that it is the superposition of two components: a
  persistent thermal emission from the wind of the B hypergiant mass
  donor \wr, and an episodic emission, perhaps a weak jet, that
  appears at the time of the X-ray outburst;

\item Further observations using the newly installed Compact Array
  Broadband Backend (CABB) will greatly improve both the estimate of
  the spectral index and the information about the polarisation of the
  radio emission of \bpcru. 

\end{itemize}

\begin{acknowledgements}
We thank the referee, Roland Walter, for his comments that clearly
improved the quality of our paper. M. P. and U. T. thank the
Swedish National Research Council for travel support to allow
observations in situ at ATCA. M. P. thanks the staff at ATCA for their
support during the stay in Narrabri, during the sessions of
remote observing and for a generous time
allocation. M. P. is supported by a two-year postdoc from the 
University of Gothenburg (Sweden). Swift/BAT transient monitor results
provided by the Swift/BAT team. This research has made use of the
SIMBAD data base operated at CDS, Strasbourg France, and NASA's
Astrophysics Data System.
\end{acknowledgements}

\bibliography{varia,starformation+IR,tech,theory,binaries_xraytrans}
\bibliographystyle{aa}

%\begin{thebibliography}{}
%\end{thebibliography}

\end{document}